\documentclass[12pt]{article}
\newcommand{\nc}{\newcommand}
\nc{\renc}{\renewcommand}

\usepackage{calc}
\usepackage{ifthen}
\usepackage{graphicx}
\usepackage{epsfig}
\usepackage{miniplot}

\usepackage{wrapfig}
\usepackage{subfigure}
\usepackage{rotfloat}

\usepackage{epsfig}
\usepackage{graphicx}
\usepackage{epsf}

\nc{\half}{{\textstyle{1\over2}}}
\nc{\etal}{\mbox{\it et al. }}
\nc{\ie}{{\it i.e.}}
\nc{\eg}{{\it e.g.}}

\renc{\thefootnote}{\arabic{footnote}}
\nc{\capt}[1]{{\bf Figure.} {\small\sl #1}}

\nc{\eqs}[2]{\mbox{Eqs.~(\ref{#1},\,\ref{#2})}}
\nc{\eq}[1]{\mbox{Eq.~(\ref{#1})}}

\nc{\figs}[2]{\mbox{Figs.~(\ref{#1},\,\ref{#2})}}
\nc{\fig}[1]{\mbox{Fig~.(\ref{#1})}}

\nc{\tag}[1]{\label{#1} \marginpar{{\footnotesize #1}}}
\nc{\mtag}[1]{\label{#1} \mbox{\marginpar{{\footnotesize #1}}}}
\renc{\baselinestretch}{1.5}
\newlength{\overeqskip}
\newlength{\undereqskip}
\nc{\be}[1]{\begin{equation} \mbox{$\label{#1}$}}
\nc{\bea}[1]{\begin{eqnarray} \mbox{$\label{#1}$}}
\nc{\Section}[2]{\section{#2}\label{#1}}
\nc{\Bibitem}[1]{\bibitem{#1}}
\nc{\Label}[1]{\label{#1}}

\nc{\eea}{\vspace{\undereqskip}\end{eqnarray}}
\nc{\ee}{\vspace{\undereqskip}\end{equation}}
\nc{\bdm}{\begin{displaymath}}
\nc{\edm}{\end{displaymath}}
\nc{\dpsty}{\displaystyle}
\nc{\bc}{\begin{center}}
\nc{\ec}{\end{center}}
\nc{\ba}{\begin{array}}
\nc{\ea}{\end{array}}
\nc{\bab}{\begin{abstract}}
\nc{\eab}{\end{abstract}}
\nc{\btab}{\begin{tabular}}
\nc{\etab}{\end{tabular}}
\nc{\bit}{\begin{itemize}}
\nc{\eit}{\end{itemize}}
\nc{\ben}{\begin{enumerate}}
\nc{\een}{\end{enumerate}}
\nc{\bfig}{\begin{figure}}
\nc{\efig}{\end{figure}}
\nc{\arreq}{&\!=\!&}
\nc{\arrmi}{&\!-\!&}
\nc{\arrpl}{&\!+\!&}
\nc{\arrap}{&\!\!\!\approx\!\!\!&}
\nc{\non}{\nonumber\\*}
\nc{\align}{\!\!\!\!\!\!\!\!&&}

\def\lsim{\; \raise0.3ex\hbox{$<$\kern-0.75em
      \raise-1.1ex\hbox{$\sim$}}\; }
\def\gsim{\; \raise0.3ex\hbox{$>$\kern-0.75em
      \raise-1.1ex\hbox{$\sim$}}\; }
\nc{\DOT}{\hspace{-0.08in}{\bf .}\hspace{0.1in}}
\nc{\Laada}{\hbox {$\sqcap$ \kern -1em $\sqcup$}}
\nc\loota{{\scriptstyle\sqcap\kern-0.55em\hbox{$\scriptstyle\sqcup$}}}
\nc\Loota{{\sqcap\kern-0.65em\hbox{$\sqcup$}}}
\nc\laada{\Loota}
\nc{\qed}{\hskip 3em \hbox{\BOX} \vskip 2ex}

\nc{\real}{{\rm I \! R}}
\nc{\Z}{{\sf Z \!\!\! Z}}
\nc{\complex}{{\rm C\!\!\! {\sf I}\,\,}}
\def\bigid{\leavevmode\hbox{\small1\kern-3.8pt\normalsize1}}
\def\id{\leavevmode\hbox{\small1\kern-3.3pt\normalsize1}}
\nc{\slask}{\!\!\!/}
\nc{\bis}{{\prime\prime}}
\nc{\pa}{\partial}
\nc{\na}{\nabla}
\nc{\ra}{\rangle}
\nc{\la}{\langle}
\nc{\goto}{\rightarrow}
\nc{\swap}{\leftrightarrow}

\nc{\EE}[1]{ \mbox{$\cdot10^{#1}$} }
\nc{\abs}[1]{\left|#1\right|}
\nc{\at}[2]{\left.#1\right|_{#2}}
\nc{\norm}[1]{\|#1\|}
\nc{\abscut}[2]{\Abs{#1}_{\scriptscriptstyle#2}}
\nc{\vek}[1]{{\rm\bf #1}}
\nc{\integral}[2]{\int\limits_{#1}^{#2}}
\nc{\inv}[1]{\frac{1}{#1}}
\nc{\dd}[2]{{{\partial #1}\over{\partial #2}}}
\nc{\ddd}[2]{{{{\partial}^2 #1}\over{\partial {#2}^2}}}
\nc{\dddd}[3]{{{{\partial}^2 #1}\over
        {\partial #2 \partial #3}}}
\nc{\dder}[2]{{{d #1}\over{d #2}}}
\nc{\ddder}[2]{{{d^2 #1}\over{d {#2}^2}}}
\nc{\dddder}[3]{{d^2 #1}\over
        {d #2 d #3}}
\nc{\dx}[1]{d\,^{#1}x}
\nc{\dy}[1]{d\,^{#1}y}
\nc{\dz}[1]{d\,^{#1}z}
\nc{\dl}[1]{\frac{d\,^{#1}l}{(2\pi)^{#1}}}
\nc{\dk}[1]{\frac{d\,^{#1}k}{(2\pi)^{#1}}}
\nc{\dq}[1]{\frac{d\,^{#1}q}{(2\pi)^{#1}}}

\nc{\cc}{\mbox{$c.c.$ }}
\nc{\hc}{\mbox{$h.c.$ }}
\nc{\cf}{cf.\ }
\nc{\erfc}{{\rm erfc}}
\nc{\Tr}{{\rm Tr\,}}
\nc{\tr}{{\rm tr\,}}
\nc{\pol}{{\rm pol}}
\nc{\sign}{{\rm sign}}
\nc{\bfT}{{\bf T }}

\def\GeV{{\rm\ GeV}}

\nc{\cA}{{\cal A}}
\nc{\cB}{{\cal B}}
\nc{\cD}{{\cal D}}
\nc{\cE}{{\cal E}}
\nc{\cG}{{\cal G}}
\nc{\cH}{{\cal H}}
\nc{\cL}{{\cal L}}
\nc{\cO}{{\cal O}}
\nc{\cT}{{\cal T}}
\nc{\cN}{{\cal N}}
\nc{\rvac}[1]{|{\cal O}#1\rangle}
\nc{\lvac}[1]{\langle{\cal O}#1|}
\nc{\rvacb}[1]{|{\cal O}_\beta #1\rangle}
\nc{\lvacb}[1]{\langle{\cal O}_\beta #1 |}
\nc{\bb}{\bar{\beta}}
\nc{\bt}{\tilde{\beta}}
\nc{\ctH}{\tilde{\cal H}}
\nc{\chH}{\hat{\cal H}}
\nc{\1}{\aa}
\nc{\2}{\"{a}}
\nc{\3}{\"{o}}
\nc{\4}{\AA}
\nc{\5}{\"{A}}
\nc{\6}{\"{O}}
\nc{\al}{\alpha}
\nc{\g}{\gamma}
\nc{\Del}{\Delta}
\nc{\e}{\epsilon}
\nc{\eps}{\epsilon}
\nc{\lam}{\lambda}
\nc{\om}{\omega}
\nc{\Om}{\Omega}
\nc{\ve}{\varepsilon}
\nc{\mn}{{\mu\nu}}
\nc{\vp}{\varphi}

\nc{\advp}[3]{{\it  Adv.\ in\ Phys.\ }{{\bf #1} {(#2)} {#3}}}
\nc{\annp}[3]{{\it  Ann.\ Phys.\ (N.Y.)\ }{{\bf #1} {(#2)} {#3}}}
\nc{\apl}[3]{{\it  Appl. Phys. Lett. }{{\bf #1} {(#2)} {#3}}}
\nc{\apj}[3]{{\it  Ap.\ J.\ }{{\bf #1} {(#2)} {#3}}}
\nc{\apjl}[3]{{\it  Ap.\ J.\ Lett.\ }{{\bf #1} {(#2)} {#3}}}
\nc{\app}[3]{{\it Astropart.\ Phys.\ }{{\bf #1} {(#2)} {#3}}}
\nc{\cmp}[3]{{\it  Comm.\ Math.\ Phys.\ }{{ \bf #1} {(#2)} {#3}}}
\nc{\cqg}[3]{{\it  Class.\ Quant.\ Grav.\ }{{\bf #1} {(#2)} {#3}}}
\nc{\epl}[3]{{\it  Europhys.\ Lett.\ }{{\bf #1} {(#2)} {#3}}}
\nc{\ijmp}[3]{{\it Int.\ J.\ Mod.\ Phys.\ }{{\bf #1} {(#2)} {#3}}}
\nc{\ijtp}[3]{{\it Int.\ J.\ Theor.\ Phys.\ }{{\bf #1} {(#2)} {#3}}}
\nc{\jmp}[3]{{\it  J.\ Math.\ Phys.\ }{{ \bf #1} {(#2)} {#3}}}
\nc{\jpa}[3]{{\it  J.\ Phys.\ A\ }{{\bf #1} {(#2)} {#3}}}
\nc{\jpc}[3]{{\it  J.\ Phys.\ C\ }{{\bf #1} {(#2)} {#3}}}
\nc{\jap}[3]{{\it J.\ Appl.\ Phys.\ }{{\bf #1} {(#2)} {#3}}}
\nc{\jpsj}[3]{{\it J.\ Phys.\ Soc.\ Japan\ }{{\bf #1} {(#2)} {#3}}}
\nc{\lmp}[3]{{\it Lett.\ Math.\ Phys.\ }{{\bf #1} {(#2)} {#3}}}
\nc{\mpl}[3]{{\it  Mod.\ Phys.\ Lett.\ }{{\bf #1} {(#2)} {#3}}}
\nc{\ncim}[3]{{\it  Nuov.\ Cim.\ }{{\bf #1} {(#2)} {#3}}}
\nc{\np}[3]{{\it  Nucl.\ Phys.\ }{{\bf #1} {(#2)} {#3}}}
\nc{\npps}[3]{{\it  Nucl.\ Phys.\ Proc.\ Suppl.\ }{{\bf #1} {(#2)} {#3}}}
\nc{\pr}[3]{{\it Phys.\ Rev.\ }{{\bf #1} {(#2)} {#3}}}
\nc{\pra}[3]{{\it  Phys.\ Rev.\ A\ }{{\bf #1} {(#2)} {#3}}}
\nc{\prb}[3]{{\it  Phys.\ Rev.\ B\ }{{{\bf #1} {(#2)} {#3}}}}
\nc{\prc}[3]{{\it  Phys.\ Rev.\ C\ }{{\bf #1} {(#2)} {#3}}}
\nc{\prd}[3]{{\it  Phys.\ Rev.\ D\ }{{\bf #1} {(#2)} {#3}}}
\nc{\prl}[3]{{\it Phys.\ Rev.\ Lett.\ }{{\bf #1} {(#2)} {#3}}}
\nc{\pl}[3]{{\it  Phys.\ Lett.\ }{{\bf #1} {(#2)} {#3}}}
\nc{\prep}[3]{{\it Phys.\ Rep.\ }{{\bf #1} {(#2)} {#3}}}
\nc{\prsl}[3]{{\it Proc.\ R.\ Soc.\ London\ }{{\bf #1} {(#2)} {#3}}}
\nc{\ptp}[3]{{\it  Prog.\ Theor.\ Phys.\ }{{\bf #1} {(#2)} {#3}}}
\nc{\ptps}[3]{{\it  Prog\ Theor.\ Phys.\ suppl.\ }{{\bf #1} {(#2)} {#3}}}
\nc{\physa}[3]{{\it  Physica\ A\ }{{\bf #1} {(#2)} {#3}}}
\nc{\physb}[3]{{\it  Physica\ B\ }{{\bf #1} {(#2)} {#3}}}
\nc{\phys}[3]{{\it Physica\ }{{\bf #1} {(#2)} {#3}}}
\nc{\rmp}[3]{{\it  Rev.\ Mod.\ Phys.\ }{{\bf #1} {(#2)} {#3}}}
\nc{\rpp}[3]{{\it Rep.\ Prog.\ Phys.\ }{{\bf #1} {(#2)} {#3}}}
\nc{\sjnp}[3]{{\it Sov.\ J.\ Nucl.\ Phys.\ }{{\bf #1} {(#2)} {#3}}}
\nc{\spjetp}[3]{{\it Sov.\ Phys.\ JETP\ }{{\bf #1} {(#2)} {#3}}}
\nc{\yf}[3]{{\it Yad.\ Fiz.\ }{{\bf #1} {(#2)} {#3}}}
\nc{\zetp}[3]{{\it Zh.\ Eksp.\ Teor.\ Fiz.\  }{{\bf #1}  {(#2)} {#3}}}
\nc{\zp}[3]{{\it Z.\ Phys.\ }{{\bf #1} {(#2)} {#3}}}
\nc{\ibid}[3]{{\sl ibid.\ }{{\bf #1} {#2} {#3}}}
\nc{\rf}[1]{(\ref{#1})}
\nc{\nn}{\nonumber \\*}
\nc{\bfB}{\bf{B}}
\nc{\bfv}{\bf{v}}
\nc{\bfx}{\bf{x}}
\nc{\bfy}{\bf{y}}
\nc{\vx}{\vec{x}}
\nc{\vy}{\vec{y}}
\nc{\oB}{\overline{B}}
\nc{\oI}{\overline{I}}
\nc{\oR}{\overline{R}}
\nc{\rar}{\rightarrow}
\nc{\ti}{\times}
\nc{\slsh}{\hskip-5pt/}
\nc{\sm}{Standard~Model~}
\nc{\MP}{M_{\rm Pl}}
\nc{\tp}{t_{\rm Pl}}
\nc{\ave}{\bar{E}}

\nc{\eff}{{\rm eff}}
\nc{\kk}{\vek{k}}
\nc{\pp}{{\rm p}}
\nc{\ga}{g_{a\gamma}}
\nc{\vv}{\\}
\nc{\eee}{{\bf E}}
\nc{\bbb}{{\bf B}}
\nc{\qcd}{T_{\rm QCD}}
\nc{\G}{\rm \ G}
\def\vec#1{{\bf #1}}

\def\lae{\;^{<}_{\sim} \;} \def\gae{\; ^{>}_{\sim} \;} 

\def\ell{e^{c}LL}

\begin{document}
{\title{\vskip-2truecm{\hfill {{\small \\
	\hfill \\
	}}\vskip 1truecm}
{\LARGE  Supergravity and Two-Field Inflation Effects in Right-Handed Sneutrino-Modified D-term Inflation 
}}
{\author{
{\sc \large Chia-Min Lin$^{1}$ and John McDonald$^{2}$}\\
{\sl\small Cosmology and Astroparticle Physics Group, University of Lancaster,
Lancaster LA1 4YB, UK}
}
\maketitle
\begin{abstract}
\noindent

   We extend previous work on the minimal D-term inflation model modified by Right-Handed (RH) sneutrino fields to include additional inflaton-sector SUGRA corrections and two-field inflation effects. We show that SUGRA corrections simultaneously allow $n_{s}$ to be within 3-year WMAP limits and the cosmic string contribution to the CMB power spectrum to be less than 5$\%$. For gauge coupling $g \leq 1$, the CMB contribution from cosmic strings is predicted to be at least 1$\%$ while the spectral index is predicted to be less than 0.968 for a CMB string contribution less than 5$\%$. Treating the inflaton-RH sneutrino system as a two-field inflation model, we show that the time-dependence of the RH sneutrino field strongly modifies the single-field results for values of RH sneutrino mass $m_{\Phi} > 0.1 H$. The running spectral index is $\alpha \approx -0.0002$ when $m_{\Phi} < 0.1 H$ but increases to positive values as $m_{\Phi}/H$ increases, with $\alpha > 0.008$ for $m_{\Phi} > 1.0 H$.

\end{abstract} 
\vfil
 \footnoterule{\small $^1$c.lin3@lancaster.ac.uk, $^2$j.mcdonald@lancaster.ac.uk}   
 \newpage 
\setcounter{page}{1}                   


\section{Introduction}  

        One clue to the nature of supersymmetric (SUSY) inflation is its compatibility with supergravity (SUGRA).
It has long been known that in the case of inflation models driven by an F-term, SUGRA corrections generate a Hubble scale inflaton mass which spoils the flatness of the inflaton potential \cite{eta}. Elimination of such corrections is possible in F-term hybrid inflation \cite{fti}, but only if there is a significant suppression of the higher-order terms in the K\"ahler potential \cite{etahi}. In contrast, D-term hybrid inflation models \cite{dti} automatically evade Hubble corrections and are therefore naturally compatible with SUGRA.

          However, there are problems with D-term inflation. 
Minimal D-term inflation ends with a phase transition which produces local cosmic strings \cite{dtics}. WMAP observations \cite{wmap} bound the contribution of local cosmic strings to the cosmic microwave background (CMB) power spectrum to be less than about 10$\%$ \cite{wyman}, which results in an upper bound on the cosmic string tension which excludes minimal D-term inflation unless the gauge and superpotential couplings are suppressed \cite{endo,csb,csb2}. In addition, the latest 3-year WMAP value of the spectral index, $n_{s} = 0.958 \pm 0.016$ (1-$\sigma$) \cite{wmap}, is significantly smaller that the minimal D-term inflation prediction, $n_{s} = 0.983$. 

                   D-term inflation has two distinct regimes corresponding to large and small couplings. If we consider unsuppressed gauge and superpotential couplings then the CMB power spectrum is completely dominated by the cosmic string contribution. However, if we consider small superpotential and gauge couplings ($ \lae 10^{-5}$ and $10^{-2}$ respectively) then it is possible to suppress the cosmic string contribution to O$(10)\%$ \cite{csb,csb2}. 
(Similar but somewhat less restrictive upper bounds were obtained in \cite{battye}.) The adiabatic density perturbation from inflation in this case has $n_{s} = 1$. However, the addition of the cosmic string contribution can lower the apparent spectral index of the CMB sufficiently for it to be consistent with the observed value \cite{battye,aur,bevis}. 
This may allow minimal D-term inflation to be consistent with the observed CMB, but at the cost of small couplings \footnote{A alternative approach has recently been suggested in \cite{bj}.}.  In the following we will consider the case of unsuppressed gauge and superpotential couplings. In this case, if minimal D-term inflation is correct, it should be modified to reduce both the string tension and the spectral index.

             How can this be achieved within a well-motivated framework? A possible solution was proposed in \cite{dlin1}. In general, the minimal D-term inflation model should be considered as a sector of a complete theory which also includes the Minimal SUSY Standard Model (MSSM), extended to include neutrino masses ($\nu$MSSM). In this case, modification of the minimal D-term inflation model could come from corrections due to $\nu$MSSM sector fields. In the context of see-saw neutrino masses, it was shown in \cite{dlin1} that a SUGRA correction due to the F-term energy density of a RH sneutrino field can lower both the string tension and spectral index. In this model, RH Sneutrino-Modified (RHSM) D-term inflation, it was found that if the RH sneutrino field is in the range (0.1-1)$M$ ($M = M_{Pl}/\sqrt{8 \pi}$) and has mass of the order of (0.1-1)$H_{I}$, 
where $H_{I}$ is the Hubble parameter during inflation, then it is possible to satisfy the CMB upper bound on the string tension for values of the spectral index that are within the lower bound from WMAP.   
This provides the only example of a minimal D-term hybrid inflation model with unsuppressed couplings that is consistent with the observed CMB. 
However, the analysis of the RHSM D-term inflation model given in \cite{dlin1} was restricted to the case where inflaton-sector SUGRA corrections are negligible and the slow-rolling RH sneutrino field can be treated as a constant. The purpose of this paper is to generalise this analysis. 
 
     SUGRA corrections to minimal D-term inflation without RH sneutrino modification were considered in \cite{osyok}. Although these allowed a reduction of both the string tension and the spectral index, it was shown that such corrections cannot lower the string tension sufficiently to be consistent with the observed CMB \cite{csb2}. Here we will show that once RH sneutrino modification is included, such corrections can substantially increase the range of spectral index for which RHSM D-term inflation is consistent with the observed CMB.

     The RHSM D-term inflation model also provides a well-motivated example of a two-field inflation model. In order to study the effects of two-field inflation we will apply the $\delta N$ method \cite{deltan,deltan2}. We will show how such effects modify the string tension and spectral index and determine the conditions under which the RH sneutrino field may be treated as constant.  

         The paper is organised as follows. In Section 2 we review the original RHSM D-term inflation model and update WMAP and cosmic string bounds to take account of recent developments. 
In Section 3 we consider the effect of combining SUGRA corrections from the inflaton sector fields with the RH sneutrino correction. In Section 4 we apply the $\delta N$ method to study two-field inflation effects of the inflaton-RH sneutrino system and establish the conditions under which the constant RH sneutrino field approximation is valid. We also compute the running spectral index due to the evolution of the RH sneutrino field.  In Section 5 we present our conclusions.

\section{RH Sneutrino-Modified D-term Inflation}

         The superpotential of the RHSM D-term inflation model is given by \cite{dlin1}  
\be{a1} W = W_{D} + W_{\nu}                ~.\ee
$W_{D}$ is the superpotential of minimal D-term hybrid inflation \cite{dti}, 
\be{e2}   W_{D} = \lambda S \Phi_{+} \Phi_{-}         ~,\ee
where $S$ is the inflaton superfield and $\Phi_{\pm}$ are superfields charged under the $U(1)_{FI}$ gauge symmetry responsible for the Fayet-Iliopoulos term. 
$W_{\nu}$ is the superpotential of the see-saw neutrino mass model \cite{seesaw}  
\be{ne1} W_{\nu} = \lambda_{\nu} \Phi H_{u} L  + \frac{m_{\Phi}}{2} \Phi^{2}   ~,\ee
where $\Phi$ is the RH neutrino superfield, $H_{u}$ and $L$ are the Higgs and lepton superfield doublets and $m_{\Phi}$ is the RH neutrino mass. In this we have suppressed generation indices. 
The scalar potential of minimal D-term inflation in global SUSY limit is
\be{e3}  V(S, \Phi_{+}, \Phi_{-}) = \lambda^{2} \left[ \left|S\right|^{2}\left(\left|\Phi_{+}\right|^{2} 
+ \left|\Phi_{-}\right|^{2} \right) + \left|\Phi_{+}\right|^{2}\left|\Phi_{-}\right|^{2} \right] 
+ \frac{g^{2}}{2} \left( \left|\Phi_{+}\right|^{2} - \left|\Phi_{-}\right|^{2} + \xi \right)^{2} 
   ~,\ee
where $\xi$ is the Fayet-Iliopoulos term. Inflation occurs when $|S| > |S|_{c} = g \xi^{1/2}/\lambda$, in which case 
$\Phi_{+}$ and $\Phi_{-}$ are equal to zero at the minimum of the potential as a function of $|S|$. The evolution of the inflaton is then determined by the 1-loop potential \cite{csb2} 
\be{e4}  V(S) = V_{o} + \frac{g^{4} \xi^{2}}{32 \pi^{2}} \left[ 2 \ln \left( \frac{|S|^{2}}{\Lambda^{2}} \right) 
+ \left(z + 1\right)^{2} \ln \left(1 + z^{-1} \right) +  \left(z - 1\right)^{2} \ln \left(1 - z^{-1} \right)
\right]    ~,\ee 
where $V_{o} = g^{2} \xi^{2}/2$ and $z = \lambda^{2} |S|^{2}/g^{2} \xi$. The effect of SUGRA, assuming a minimal K\"ahler potential, is to replace $|S|^{2}$ by $|S|^{2}e^{|S|^{2}/M^{2}}$, where $M = M_{Pl}/\sqrt{8 \pi}$ \cite{csb2}. At $|S| = |S|_{c}$ there is a $U(1)$-breaking phase transition, ending inflation and producing local cosmic strings.  

        In \cite{dlin1} a K\"ahler potential consisting of minimal kinetic terms and an interaction between $S$ and $\Phi$ was considered, 
\be{e2a}   K  =  S^{\dagger}S + \Phi^{\dagger}\Phi + \frac{c S^{\dagger}S \Phi^{\dagger} \Phi}{M^{2}}       ~.\ee
In the limit where $|S|^{2} \ll M^{2}$ and $z \gg 1$, which requires that $g \lae 0.1$, the SUGRA inflaton potential 
is well-approximated by the global SUSY potential,
\be{e4}  V(S) = V_{o} +  \frac{g^{4} \xi^{2}}{16 \pi^{2}} \ln \left( \frac{|S|^{2}}{\Lambda^{2}} \right)   ~.\ee
For $|\Phi|^2/M^2 \ll 1$, the RH sneutrino potential is given by the global SUSY F-term, 
\be{e4a} V(\Phi) = |F_{\Phi}|^{2} = m_{\Phi}^{2} |\Phi|^{2} ~.\ee
Expanding the SUGRA potential in $|S|^2/M^2$, the leading-order 
effect of $c \neq 0$ is to introduce a potential interaction between $S$ and $\Phi$ \cite{dlin1},  
\be{e7}   \Delta V  =  - \frac{ \left(c-1\right) m_{\Phi}^{2} |\Phi|^{2} |S|^{2}}{M^{2}}  ~.\ee
The scalar potential in the limit $g \lae 0.1$ is therefore given by \cite{dlin1} 
\be{e10} V(s,\phi) = V_{o} + \frac{\alpha}{2} \ln \left( \frac{s^{2}}{2 \Lambda^{2}} \right)  - \frac{\kappa s^{2} \phi^{2}}{4} 
+ \frac{m_{\Phi}^{2} \phi^{2}}{2}    ~,\ee
($s = \sqrt{2} Re(S)$, $\phi = \sqrt{2} Re(\Phi)$) where\footnote{The definition of $\kappa$ used here differs from that in \cite{dlin1} by a factor $|\Phi|^2$.} 
$$    \alpha = \frac{g^{4} \xi^{2}}{8 \pi^{2}} \;\;,\;\;\;\; \kappa = \frac{ \left(c-1\right)m_{\Phi}^{2} }{M^{2}}           ~.$$
In general \eq{e10} describes a two-field inflation model. In \cite{dlin1} it was assumed that the slow-rolling $\Phi$ field could be considered constant if $m_{\Phi}$ is sufficiently small compared to $H$, in which case the standard single-field inflation analysis of density perturbations can be applied. 
In this case \eq{e7} is effectively a negative inflaton mass squared when $c > 1$. The solutions for the slow-rolling inflaton field, spectral index and curvature perturbation can then be calculated analytically\footnote{The form of the potential in this limit has also been analysed in \cite{hilltop} and \cite{hilltop2} as 'hilltop inflation'.}:  
\be{e12}   s^{2}(t) = s^{2}_{o} \left(1 - e^{-\gamma} 
\right)      ~,\ee
\be{e14} n_{s} = 1 - \frac{\gamma}{N} \left(\frac{1}{\left(1 - e^{-\gamma}\right)} + 1 \right)  ~,\ee
and  
\be{e17} P_{\zeta} =  \frac{N \xi^{2}}{3 M^{4}} \frac{1}{\Gamma}  \;\;\;;\;\;\;\; 
 \Gamma  = \frac{\gamma e^{-2 \gamma} } { \left( 1  - e^{-\gamma} \right)}  ~,\ee 
where  
\be{e18} 
 \gamma = \kappa N \phi^{2}/3 H^{2}     ~.\ee
The value of $\xi^{1/2}$ required to account for the observed curvature perturbation, 
$P_{\zeta}^{1/2} = 4.8 \times 10^{-5}$, is then
\be{e20}    \xi^{1/2} = 7.9 \times 10^{15} \Gamma^{1/4} \left(\frac{60}{N}\right)^{1/4} \GeV      ~.\ee
These solutions only depend on a single parameter, $\gamma$. As a result, it is possible to eliminate $\gamma$ and 
express $\xi^{1/2}$ as a function of $n_{s}$, allowing the model to be directly compared with CMB bounds on the cosmic string tension and spectral index \cite{dlin1}.

\subsection{Updated bounds on RHSM D-term inflation}

     Since \cite{dlin1} was published there have been developments in both the 3-year WMAP spectral index 
bound \cite{wmap} and the CMB upper bound on the cosmic string tension \cite{bevis}:
\newline $\bullet$ In the latest version of the 3-year WMAP analysis \cite{wmap}, the 
1-$\sigma$ mean value of the spectral index using 'WMAP only' data is given as $n_{s} = 0.958 \pm 0.016$, compared with the 
first version value $n_{s} = 0.951^{+0.015}_{-0.019}$. The 1-$\sigma$ mean value from 'WMAP + ALL' other data sets is given as $n_{s} = 0.947 \pm 0.015$. We will compare our results with both of these ranges. 
\newline $\bullet$ In \cite{bevis} a new value was obtained for the $l = 10$ WMAP normalised cosmic string tension $\mu$, $G \mu = 2.0 \times 10^{-6}$, based on a field-theory simulation of the cosmic string network. (The cosmic string tension in D-term inflation is given by $\mu = 2 \pi \xi$ \cite{dtics}.) We will compare our model with this new value.

      In Figures 1 and 2 we display upper limits of 10$\%$ and $5\%$ on the contribution of local cosmic strings to the CMB, based on the normalised $\mu$ of \cite{bevis}. In this case the upper bounds on $\xi^{1/2}$ are $3.9 \times 10^{15} \GeV$ (10$\%$) and $3.3 \times 10^{15} \GeV$ (5$\%$). In \cite{dlin1} an earlier 10$\%$ upper bound from \cite{endo} was used, $\xi^{1/2} < 4.6 \times 10^{15} \GeV$. In the following we will consider a model to be consistent with the observed CMB if the cosmic string contribution is less than 5$\%$ and $n_{s}$ from the inflationary adiabatic spectrum is compatible with WMAP bounds \footnote{A more detailed analysis based on combining the adiabatic and cosmic string CMB power spectra 
and comparing with WMAP data will be the subject of future work.}.  

            In Figure 1 we show $\xi^{1/2}$ as a function of the adiabatic RHSM D-term inflation 
spectral index $n_s$. In Figure 1(a) 
we show the 1- and 2-$\sigma$ upper and lower bounds on $n_{s}$ based on 'WMAP only' data. In Figure 1(b) we show the 
bounds on $n_{s}$ for the case of 'WMAP + ALL' data sets. 
In the case of 'WMAP only' data, the spectral index is consistent with the 10$\%$ cosmic string upper bound if we consider the 2-$\sigma$ lower bound on $n_{s}$, but not if we consider the 1-$\sigma$ lower bound. In the case of 'WMAP + ALL', the cosmic string upper bound is also marginally consistent with the 1-$\sigma$ lower bound. 
For the 5$\%$ cosmic string upper bound, the spectral index is generally inconsistent with 'WMAP only' data, but is compatible with the 2-$\sigma$ lower bound for 'WMAP + ALL'.

\begin{figure} 
  \centering 
  \subfigure[WMAP data only.]{ 
\label{a}
  \includegraphics[width=3.5in, angle=-90]{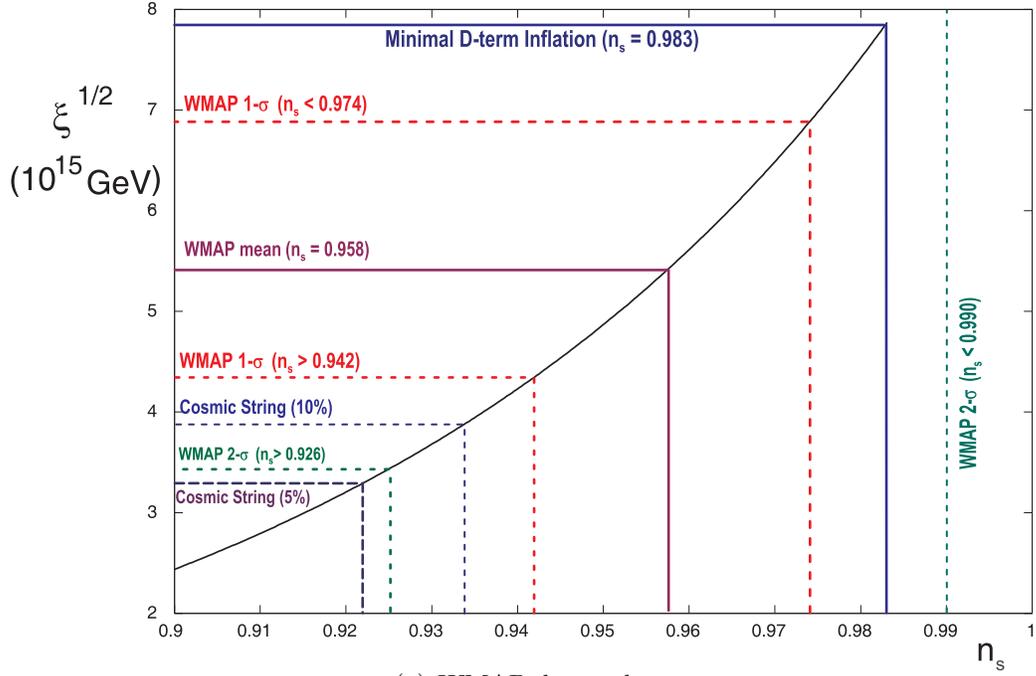} }
 \hspace{0.5in} 
  \subfigure[WMAP + ALL data.]{
\label{b}  
      \includegraphics[width=3.5in, angle=-90]{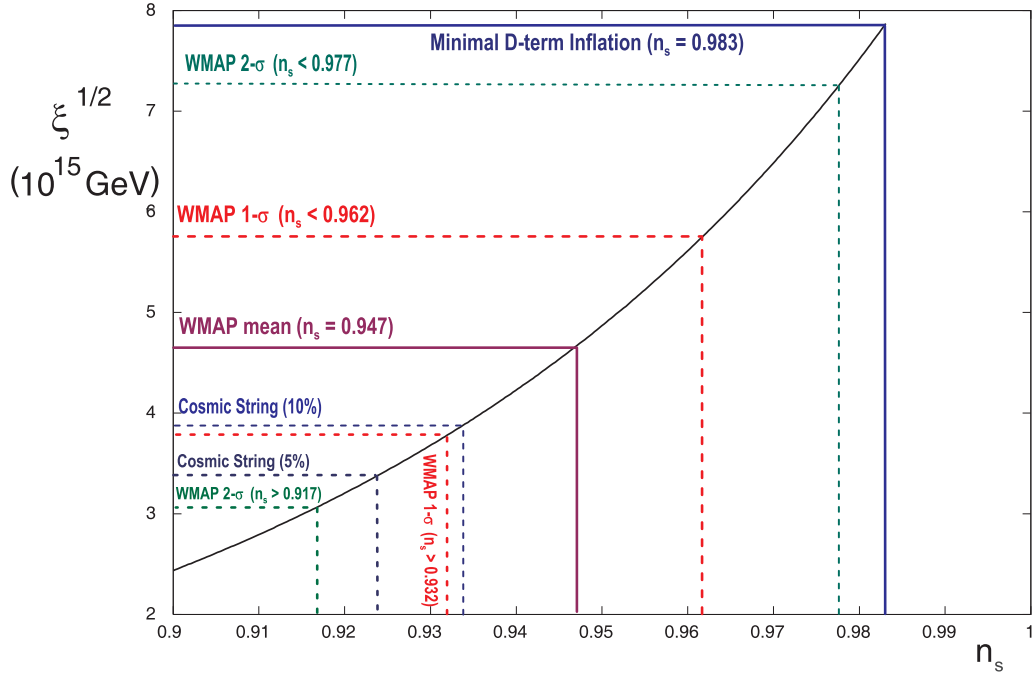}}
\caption{\footnotesize{$\xi^{1/2}$ as a function of $n_{s}$ in the original RHSM D-term Inflation model. The 1-$\sigma$ and 2-$\sigma$ bounds on $n_{s}$ are shown for the case of (a) 'WMAP only' data and (b) 'WMAP + ALL' data. The cosmic string 10$\%$ and 5$\%$ upper bounds on $\xi^{1/2}$ are also shown.}}
\end{figure}

\begin{figure}[h] 
                    \centering                   
                    \includegraphics[width=0.5\textwidth, angle=-90]{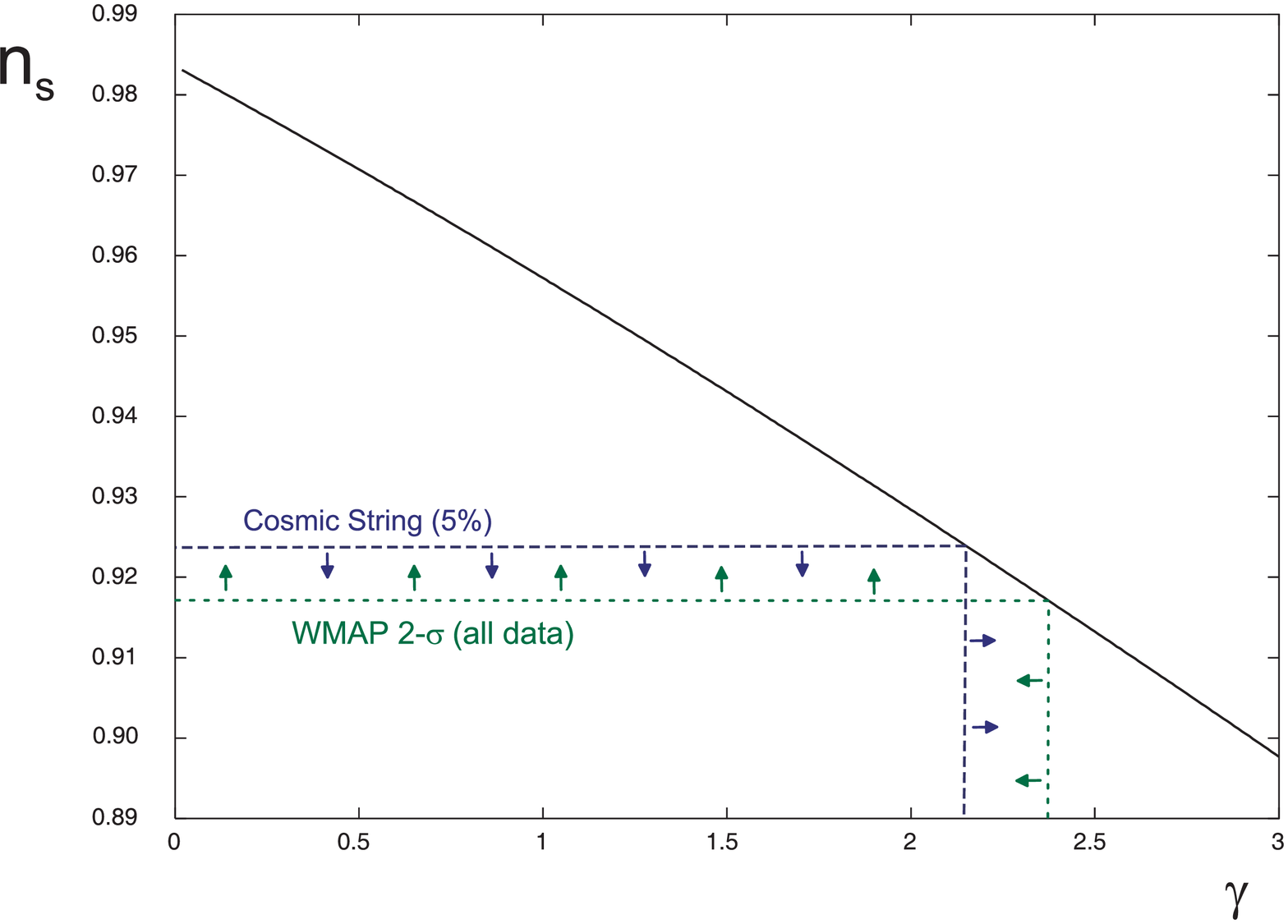}
                    \caption{\footnotesize{Spectral index as a function of $\gamma$. The 2-$\sigma$ WMAP + ALL lower bound 
                           and 5$\%$ cosmic string upper bound on $n_{s}$ are shown.}}
                    \end{figure}

            In Figure 2 we show the range of $\gamma$ as a function of $n_{s}$ required to fit the WMAP bounds. In general $\gamma$ of order 1 is necessary to significantly modify $n_{s}$ and $\xi^{1/2}$. This requires that $0.1 M \lae |\Phi| < M$ and $0.1 H \lae M_{\Phi} < H$, where the upper bounds follow from the requirements of (i) small SUGRA corrections to the $\Phi$ potential and (ii) a slow-rolling $\Phi$ field \cite{dlin1}. The 
value of $H$ during inflation is given by $H_{I} \approx g  \times 2.1 \times 10^{12} \GeV$ (using $\xi^{1/2} \approx 3.5 \times 10^{15} \GeV$), 
so with $g \approx 0.1$ the required order of magnitude of $m_{\Phi}$ for RH sneutrino modification is $10^{11} \GeV$. This is a plausible magnitude for RH neutrino masses in the see-saw neutrino mass model \cite{seesaw}.

     The RHSM D-term inflation model described above provides the only existing example of a minimal D-term inflation model with unsuppressed couplings which is compatible with WMAP observations. However, the above analysis is restricted to the case where: 
\newline (i) The only significant SUGRA correction is that coming from the non-minimal K\"ahler interaction between $S$ and $\Phi$. This will be true only if $g \lae 0.1$, so that $|S|^2/M^2 \ll 1$ and the inflaton sector SUGRA corrections are small.
\newline (ii) The RH sneutrino can be treated as constant when calculating the curvature perturbation. 
\newline In the following sections we will discuss the effect of generalising the model beyond these assumptions.

\section{Additional SUGRA Corrections} 

         In this section we consider the effect of additional SUGRA corrections due to the D-term inflation fields  
$S$, $\Phi_{+}$ and $\Phi_{-}$. We now consider the K\"ahler potential \cite{osyok,csb2}
\be{c1}
K=|S|^2+|\Phi_+|^2+|\Phi_-|^2 + \frac{c S^{\dagger}S \Phi^{\dagger} \Phi}{M^{2}} + f_+\left(\frac{|S|^2}{M^2}\right)|\Phi_+|^2+f_-\left(\frac{|S|^2}{M^2}\right)|\Phi_-|^2+b\frac{|S|^4}{M^2} 
~,\ee
with
\[
 f_\pm\left(\frac{|S|^2}{M^2}\right)=\frac{c_\pm |S|^2}{M^2} \equiv \frac{c_\pm}{2} \frac{s^2}{M^{2}}
~.\]
With $c \neq 0$ the full potential is given by adding RH sneutrino corrections to the $c = 0$ SUGRA inflaton potential derived in \cite{csb2}:    
\be{c3}
V(s,\phi)=\frac{g^2\xi^2}{2}\left\{1+\frac{g^2}{16\pi^2}\left[2\ln\left(\frac{g^2\xi z}{\Lambda^2}\right)+f_V(z)\right]\right\} - \frac{\kappa m_\phi^2s^2\phi^2}{4}+\frac{m_\phi^2\phi^2}{2}
~,\ee
where
\[
f_V(z)=(z+1)^2\ln\left(1+\frac{1}{z}\right)+(z-1)^2\ln\left(1-\frac{1}{z}\right)
\]
and
\[
z\equiv\frac{\lambda^2 s^2}{2g^2\xi}\exp\left(\frac{s^2}{2M^2}+b\frac{s^4}{4M^4}\right)\frac{1}{(1+f_+)(1+f_-)}
~.\]

\begin{figure} 
  \centering 
  \subfigure[WMAP only]{ 
    \label{wmap01:a} 
    \includegraphics[width=4.2in]{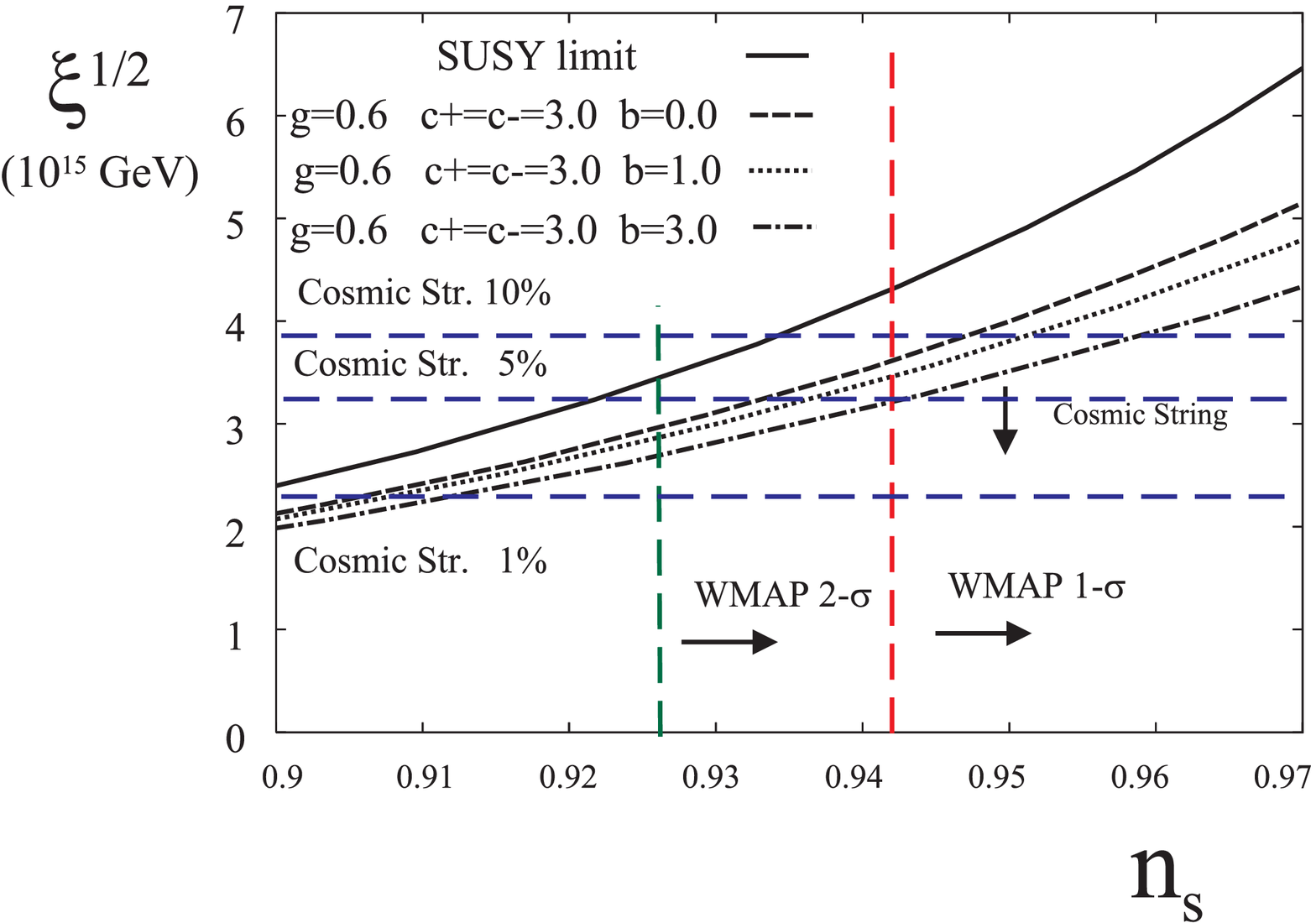}} 
  \vspace{0.3in} 
  \subfigure[WMAP + ALL data]{ 
    \label{wmap01:b} 
    \includegraphics[width=4.2in]{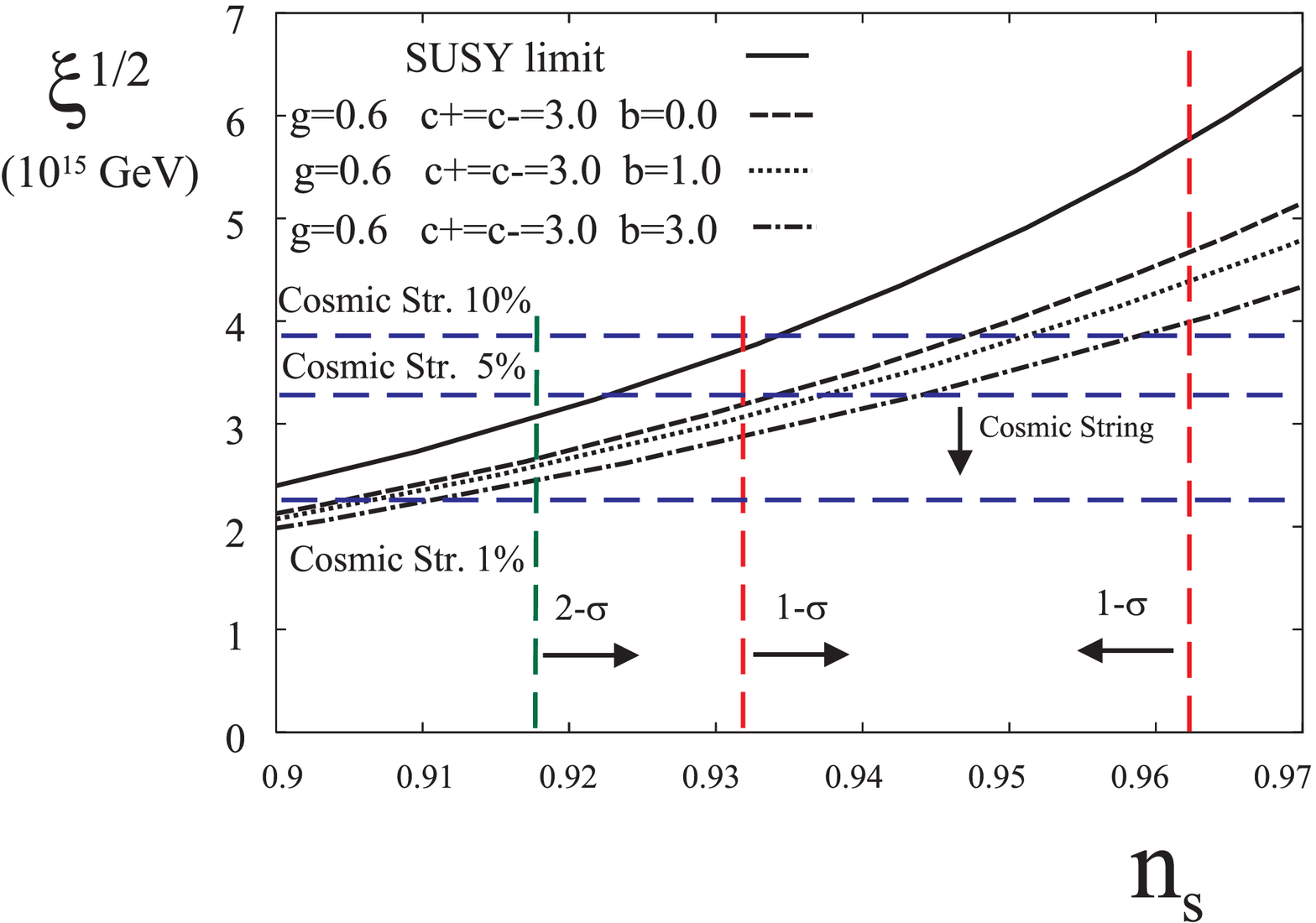}} 
  \caption{$c_+=c_-=3.0$, $g=0.6$ plot of $\xi^{1/2}$ versus $n_s$ for $b=0.0$, 1.0 and 3.0.} 
  \label{wmap01} 
\end{figure}

\begin{figure} 
  \centering 
  \subfigure[WMAP only]{ 
    \label{wmap02:a} 
    \includegraphics[width=4.2in]{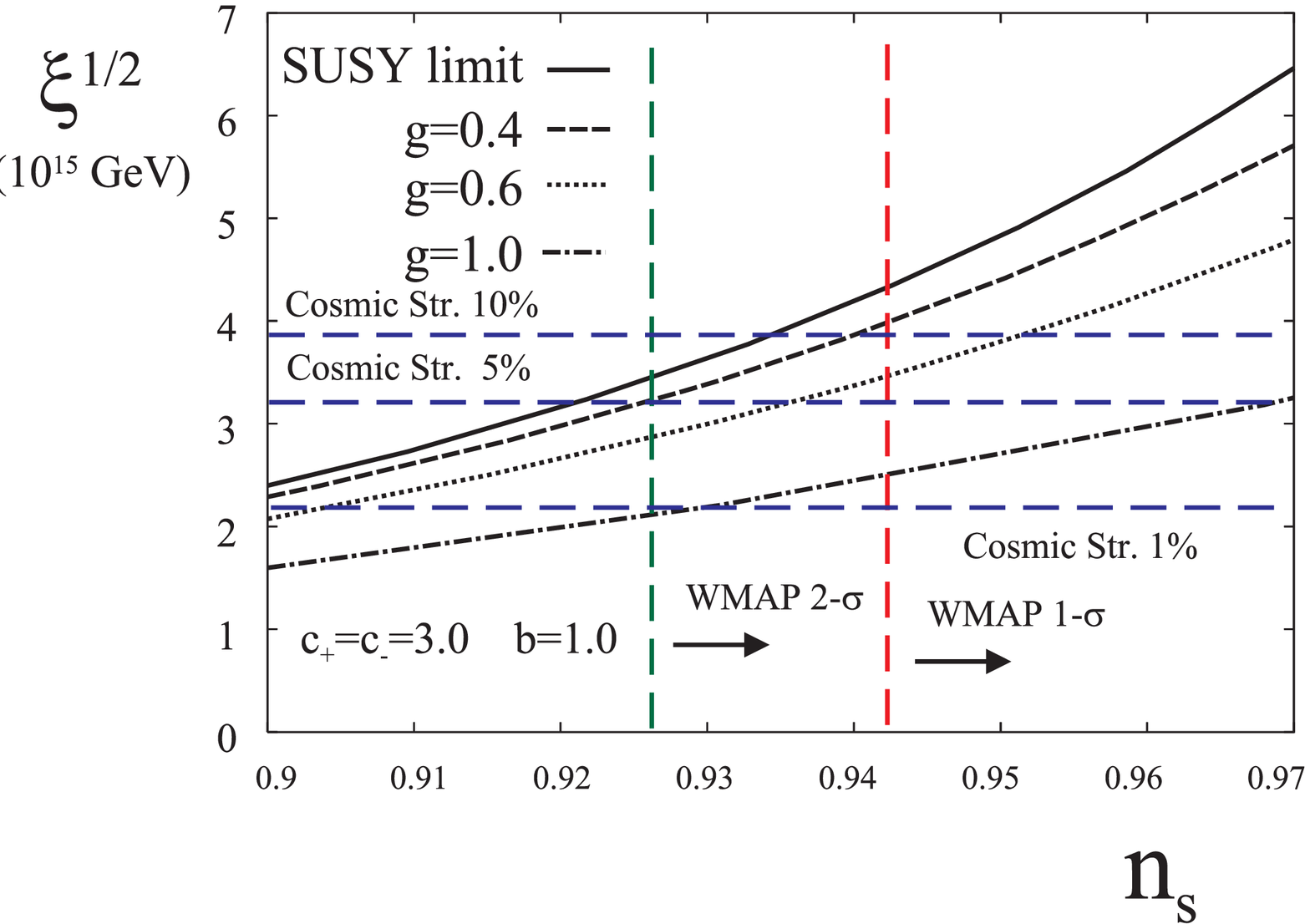}} 
  \vspace{0.3in} 
  \subfigure[WMAP + ALL data]{ 
    \label{wmap02:b} 
    \includegraphics[width=4.2in]{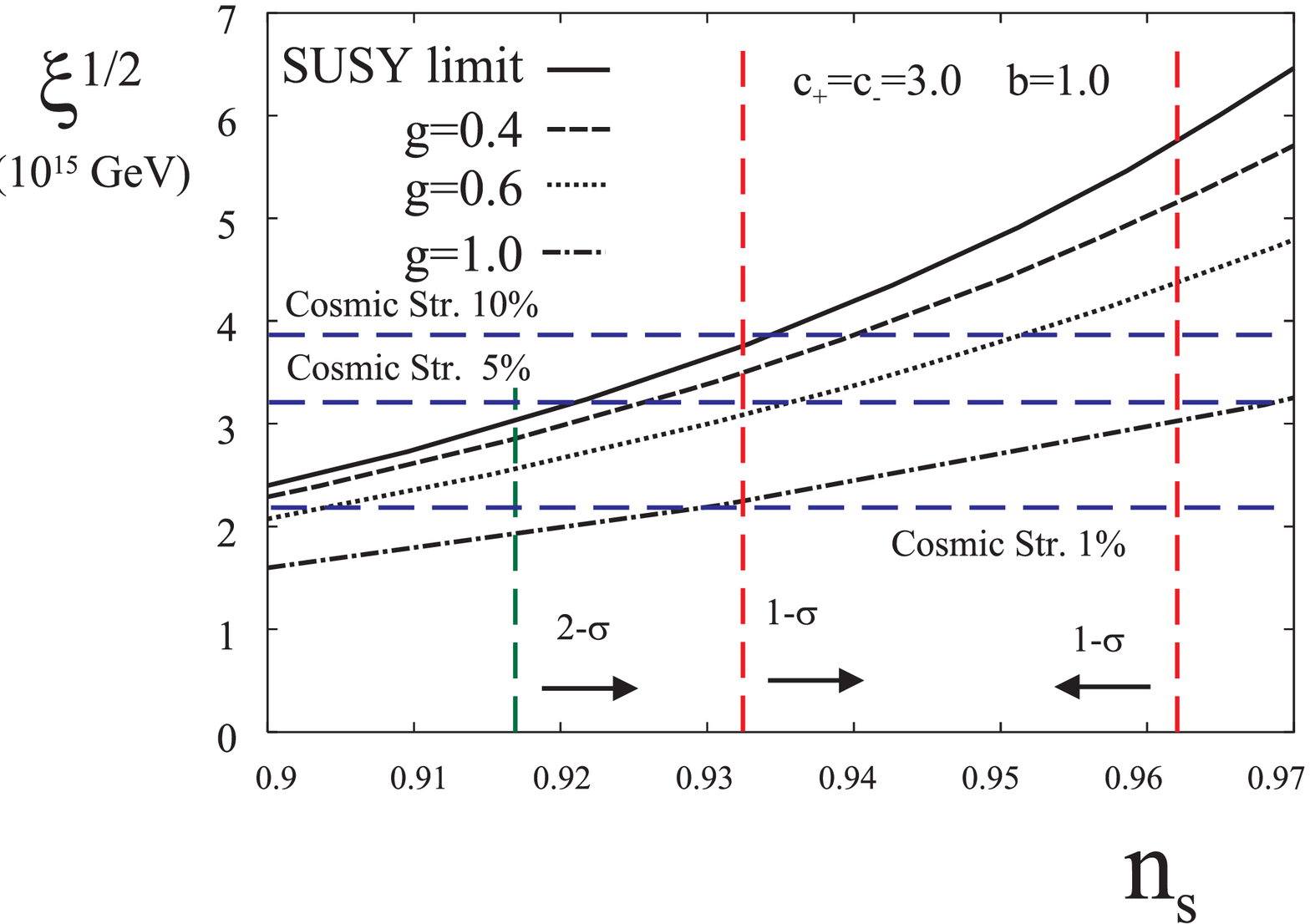}} 
  \caption{$c_+=c_{-}=3.0$, $b=1.0$ plot of $\xi^{1/2}$ versus $n_{s}$ for $g=0.4$, 0.6 and 1.0.} 
  \label{wmap02} 
\end{figure}

\begin{figure} 
  \centering 
    \includegraphics[width=4.2in]{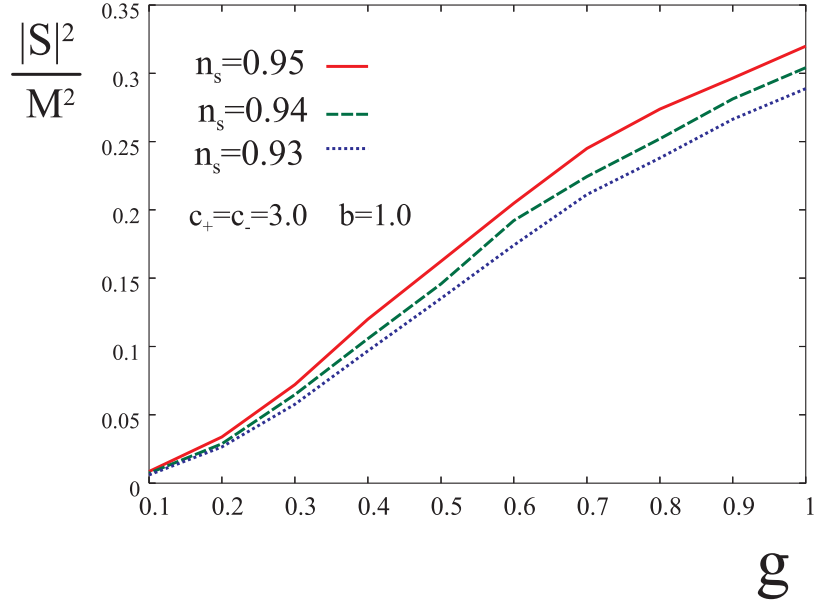} 
  \caption{$c_+=c_- = 3.0$, $b=1.0$ plot of $|S|^2/M^2$ at $N=60$ versus $g$ for $n_s =0.93$, 0.94 and 0.95.} 
  \label{wmap03} 
\end{figure}

\begin{figure} 
  \centering 
    \includegraphics[width=4.2in]{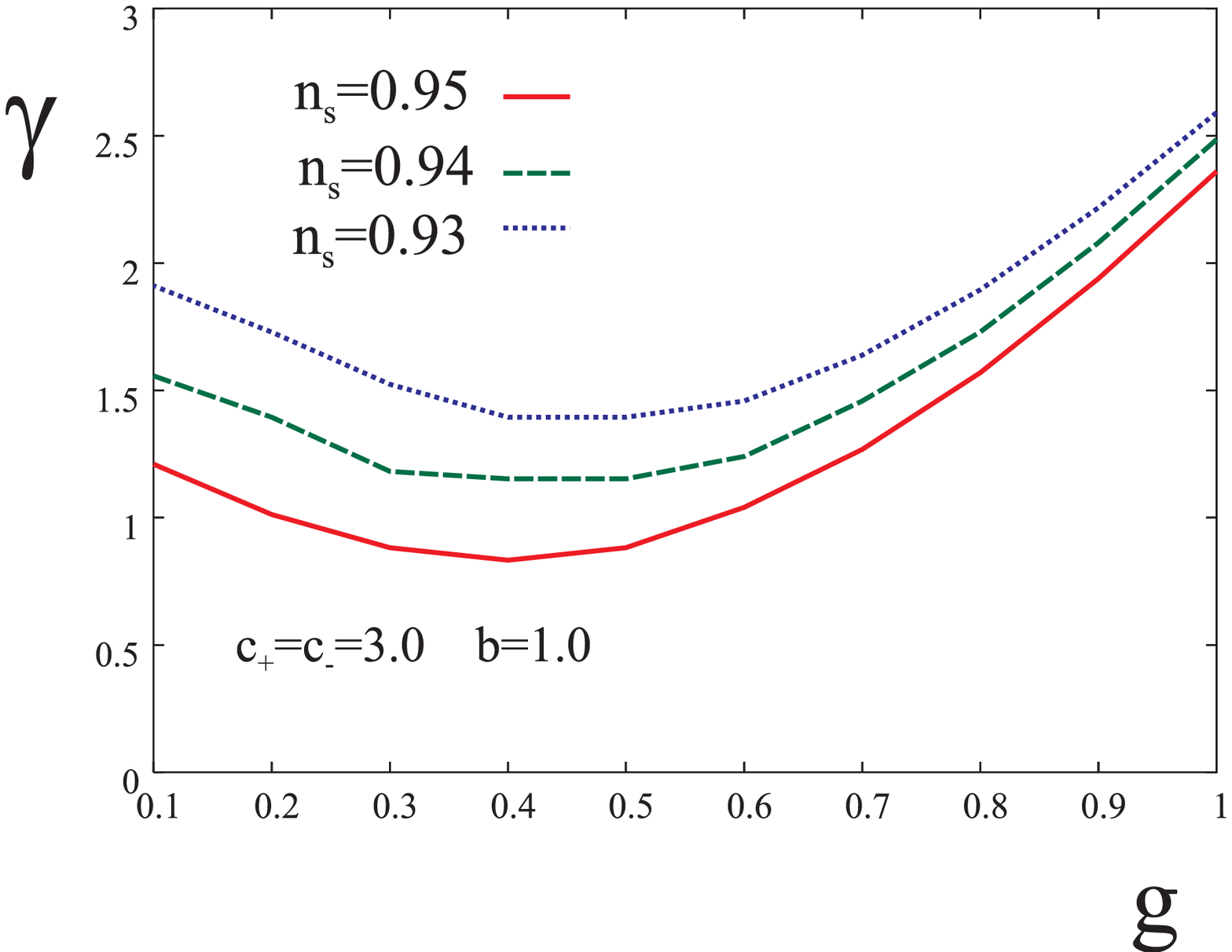} 
  \caption{$c_+=c_- = 3.0$, $b=1.0$ plot of $\gamma$ as a function of $g$ for $n_s =0.93$, 0.94 and 0.95.} 
  \label{wmap04} 
\end{figure}

       In Figures \ref{wmap01} and \ref{wmap02} we show $\xi^{1/2}$ as a function of $n_{s}$ obtained by numerically solving the $s$ field equation with constant $\phi$ and using the standard single-field inflation results for $n_{s}$ and $P_{\zeta}$ \cite{lr}. We compare these results with the most recent 'WMAP only' and 'WMAP + ALL' bounds on $n_{s}$ and with the 10$\%$, 5$\%$ and 1$\%$ cosmic string upper bound on $\xi^{1/2}$.  We fix $c = 2$ throughout.

Figure \ref{wmap01} shows results for $g = 0.6$, $c_{+} = c_{-} = 3.0$ and 
$b$ ranging from 0.0 to 3.0.  The line labelled 'SUSY limit' is the case without additional SUGRA corrections, corresponding to the original RHSM D-term inflation model with potential \eq{e10}. From Figure \ref{wmap01:a} (WMAP only data) we find that a cosmic string contribution less than 5$\%$ is compatible with the 2-$\sigma$ WMAP lower bound on $n_{s}$ for all $b > 0$ and with the 1-$\sigma$ bound for 
$b > 3.0$. From Figure \ref{wmap01:b} (WMAP + ALL data) a cosmic string contribution less than 5$\%$ is 
compatible with the 1-$\sigma$ WMAP lower bound for all $b > 0$. 

Figure \ref{wmap02} shows results for $b = 1.0$, $c_{+} = c_{-} = 3.0$ with $g$ ranging from 0.4 to 1.0. From Figure
\ref{wmap02:a} (WMAP only) we find that a cosmic string contribution below 5$\%$ is compatible with the WMAP 2-$\sigma$ lower bound on $n_{s}$ for $g > 0.4$, but would require $g > 0.6$ to be compatible at 1-$\sigma$. From Figure 
\ref{wmap02:b} (WMAP + ALL) a 5$\%$ cosmic string contribution is compatible with 2-$\sigma$ bounds for all $g$, and is compatible with 1-$\sigma$ bounds  for all $g > 0.6$. The case $g = 1$ is easily compatible with 1-$\sigma$ bounds, in which case the cosmic string contribution is in the range 1-5$\%$. The spectral index is $n_{s} \lae 0.968$ for a cosmic string contribution less than $5\%$.

    Thus combining additional SUGRA corrections with RH sneutrino modification allows the minimal D-term inflation model to be completely compatible with WMAP observations of the CMB. A sufficiently small cosmic string contribution and values of $n_{s}$ larger than the WMAP lower bounds can be obtained 
for reasonable values of $g$, $c_{\pm}$ and $b$.

          In the figures we have considered relatively large values of $g$. We must check that the value of $|S|^2/M^2$ is small enough that we can expand the K\"ahler potential in powers of $|S|^2/M^2$. In Figure \ref{wmap03} we show $|S|^2/M^2$ at $N = 60$ as a function of $g$. For the range of values of $g$ of interest, $|S|^2/M^2$ is sufficiently small that higher-order SUGRA corrections to $V(s,\phi)$ will not have a large effect.

In Figure \ref{wmap04} the value of $\gamma$ at $N = 60$ is shown as a function of $g$. Values of $\gamma$ between 1 and 2.5 are required to achieve sufficient modification of the potential.

  The cosmic string contribution is generally larger than about 1$\%$ for the parameters we have considered.  Therefore we can expect to observe the cosmic string contribution as the precision of CMB observations improves. In addition, the $5\%$ cosmic string upper bound on $\xi^{1/2}$ requires values of $n_{s}$ less than 0.968 for $g \leq 1$, with smaller values for smaller cosmic string contributions. These may be regarded as predictions of the RHSM D-term inflation model.

\section{Two-field Inflation Effects}

     RHSM D-term inflation is a two-field inflation model based on $s$ and $\phi$. Therefore the final adiabatic perturbation and spectral index will depend on the evolution throughout inflation of the adiabatic and isocurvature field, which will be linear combinations of $s$ and $\phi$ \cite{tfi}. (Field trajectories for a range of $m_{\Phi}/H$ are shown in Figure \ref{traj}.) Since the RH sneutrino field will decay well before the inflaton field, the final perturbation will be purely adiabatic. However, the adiabatic curvature perturbation will be determined by the full history of the adiabatic and isocurvature scalars from horizon exit during inflation. In this case the usual single-field expressions for the spectral index and curvature perturbation will no longer apply. The results of the previous sections are based on the assumption that $\phi$ is sufficiently slowly rolling that the model can be treated as a single-field inflation model, with the value of $\phi$ fixed to its value at horizon exit. In this section we will establish the conditions under which this 'constant $\phi$' approximation is good and  
how the results are modified by its breakdown. Since our concern here is to show how two-field inflation effects modify the single-field results, we will focus on the simpler example of the original RHSM D-term inflation model defined by the potential \eq{e10}.

\begin{figure} 
  \centering 
    \includegraphics[width=4.0in, angle = 0]{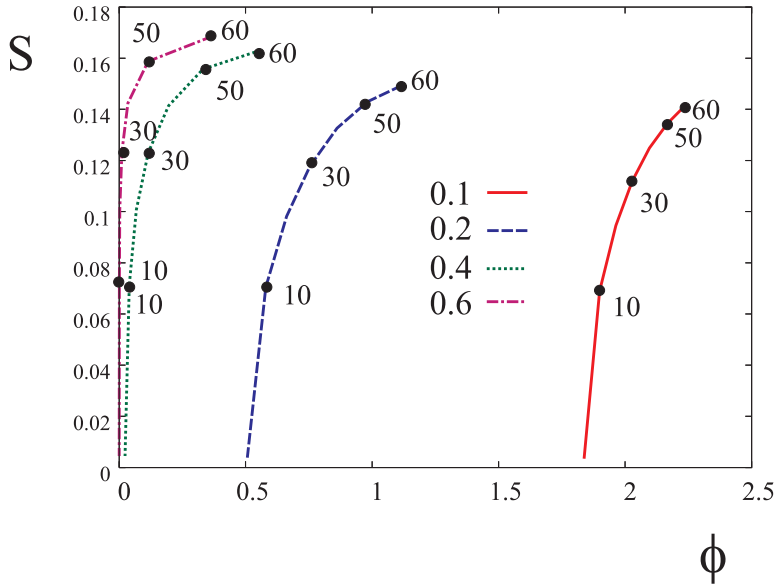} 
  \caption{Two-field inflation trajectories for $m_{\Phi}/H =$ 0.1, 0.2, 0.4 and 0.6. Values of $N$ are shown on the trajectories.} 
  \label{traj} 
\end{figure}

   To estimate the effect of two-field inflation we will apply the $\delta N$ method \cite{deltan,deltan2}. 
The curvature perturbation between superhorizon-sized patches at two different points in space is given by the difference in the number of e-foldings of inflation experienced by the two patches from an initial flat slicing where the fields have assigned values until a final slicing which has a fixed value of $\rho$. We will calculate the curvature perturbation when the density equals $\rho_{c}$, corresponding to the value of $\rho$ for the unperturbed fields when the phase transition at $s = s_{c}$ ends inflation. Although it is possible that there could be further evolution of the 
curvature perturbation and $\delta N$ after inflation, we expect that most of the effect of the RH sneutrino field on the curvature perturbation occurs during inflation. The curvature perturbation is then given by  \cite{deltan,deltan2}
\be{e19a} \zeta(t_{c},\vec{x}) = \delta N =  N(\rho_{c}, s(\vec{x}), \phi(\vec{x}))  - N(\rho_{c}, s_{i}, \phi_{i}) ~.\ee
Here $s(\vec{x}), \phi(\vec{x})$ are the perturbed initial values of the fields relative to the unperturbed initial fields $s_{i}$ and $\phi_{i}$, and $t_{c}$ is the time at which inflation ends for the unperturbed fields.

          To lowest-order, the shift in the number of e-foldings will be given by 
\be{e20} \delta N = \frac{\partial N}{\partial s_{i}} \delta s_{i} + \frac{\partial N}{\partial \phi_{i}} \delta \phi_{i}  
~,\ee
where $N \equiv N(s_{i},\phi_{i})$ is the unperturbed number of e-foldings at the end of inflation,
\be{e20a} N(s_{i},\phi_{i}) = \int_{t_{i}}^{t_{c}} H(t; s_{i},\phi_{i}) dt   ~\ee
and $\delta s_{i}$ and $\delta \phi_{i}$ are the initial quantum fluctuations at horizon exit.
The power spectrum of the curvature perturbation at the end of inflation is then  
\be{e21} P_{\zeta} = \left(\frac{\partial N}{\partial s_{i}}\right)^{2} P_{\delta s} + \left(\frac{\partial N}{\partial \phi_{i}}\right)^{2} P_{\delta \phi}      ~\ee  
where $P_{\delta s} = P_{\delta \phi} = (H/2 \pi)^2$.

We have computed the derivatives $\partial N/\partial s_{i}$ and $\partial N/\partial \phi_{i}$ by perturbing the initial conditions and solving the full $s$ and $\phi$ field equations based on \eq{e10} until $\rho$ equals the same value $\rho_{c}$ as in the unperturbed case. The initial $s_{i}$ and $\phi_{i}$ and the value of $\xi^{1/2}$ were chosen to give $N = 60$ unperturbed e-foldings, a correctly normalized $P_{\zeta}$ ($P_{\zeta}^{1/2} = 4.8 \times 10^{-5}$) and a fixed value 
for $\gamma$ calculated from \eq{e18}. This allows us to directly compare the $\delta N$ results with the constant 
$\phi$ results for $n_{s}$ and $P_{\zeta}$, which depend only on the value of $\gamma$.

       In Table 1 we give values of the derivatives of $N$ with respect to $s$ and $\phi$ as a function of $m_{\Phi}/H$. In this we have fixed $\gamma = 1$, $g = \lambda = 0.1$ and $c = 2$. From this we see that the contribution of $(\partial N/\partial \phi_{i})^{2}$ 
to $P_{\zeta}$ is negligible. Therefore quantum fluctuations of the RH sneutrino field will have a negligible effect on the 
curvature perturbation.

\begin{table}[h]
 \begin{center}
 \begin{tabular}{|c|c|c|}
	\hline   $m_{\Phi}/H$ & $\partial N/\partial s_{i}$ &  $\partial N/\partial \phi_{i}$   \\ 
	\hline   0.05 & 1443.7 & 18.8 \\
      \hline   0.2 & 970.6 & 39.1 \\
      \hline   0.4 & 757.3 & 29.3 \\
      \hline   0.6 & 717.9 & 21.5 \\
      \hline   0.8 & 705.2 & 16.6 \\
      \hline   1.0 & 699.8 & 13.4 \\
	\hline       
 \end{tabular}
 \caption{\footnotesize{$\partial N/\partial s_{i}$ and $\partial N/\partial \phi_{i}$ as a function of 
$m_{\Phi}/H$.}}  
 \end{center}
 \end{table}

     The effect of the time evolution of the classical $\phi$ field is seen by increasing the value of $m_{\Phi}/H$. 
In Figure \ref{si:a} we show the ratio $\xi^{1/2}/\xi_{o}^{1/2}$ as a function of $m_{\Phi}/H$, where $\xi_{o}^{1/2} = 7.9 \times 10^{15} \GeV$ is the value in conventional minimal D-term inflation without RH sneutrino modification. 
In Figure \ref{si:b} we show the spectral index as a function of $m_{\Phi}/H$.

\begin{figure} 
  \centering 
 \subfigure[$\xi^{1/2}$/$\xi_{o}^{1/2}$ vs. $m_{\Phi}/H$]{ 
\label{si:a} 
    \includegraphics[width=4.0in, angle = -90]{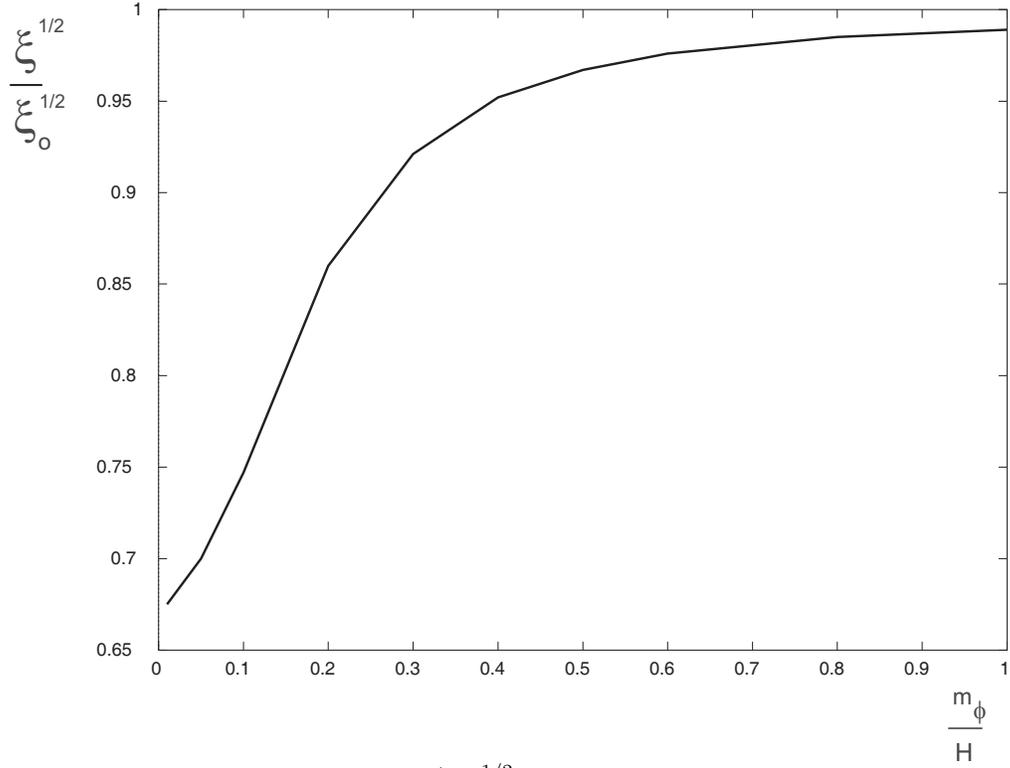}} 
   \hspace{1.5in} 
\subfigure[Spectral index vs. $m_{\Phi}/H$]{ 
    \label{si:b} 
    \includegraphics[width=4.0in, angle = -90]{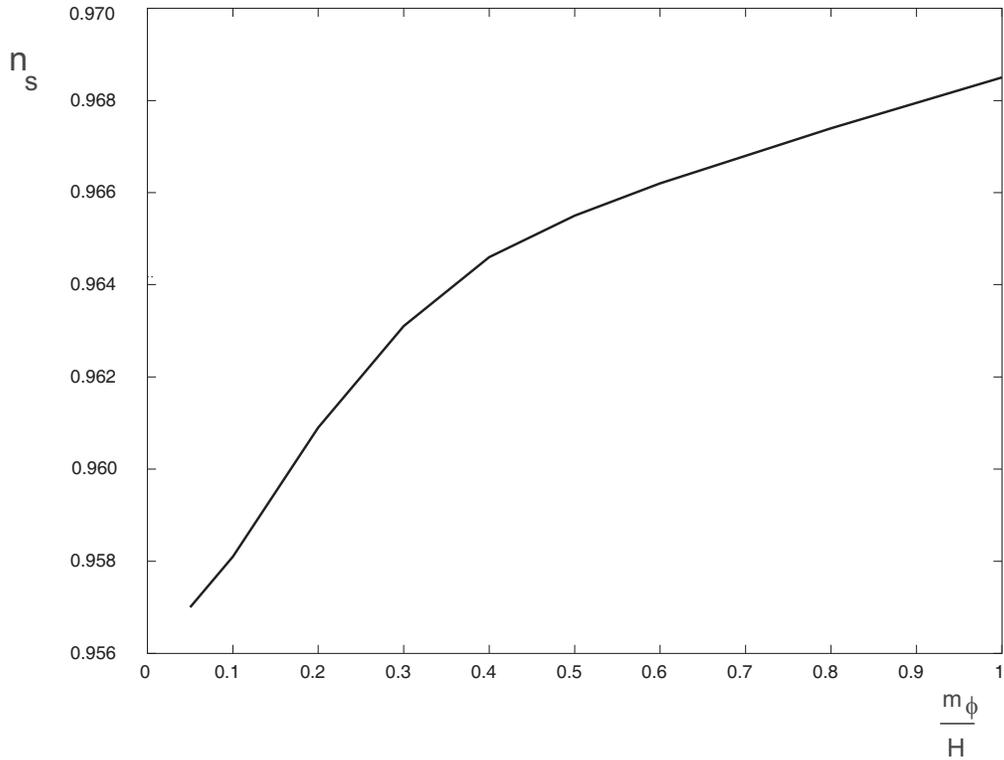}} 
\caption{Effect of $\phi$ time evolution on $\xi^{1/2}/\xi^{1/2}_{o}$ and $n_{s}$.} 
 \label{si} 
\end{figure}

                From Figure \ref{si:a} there is a very strong effect on the value of $\xi^{1/2}$ 
from the time evolution of $\phi$. For $m_{\Phi}/H = 0.2$, the suppression of $\xi^{1/2}$ 
has become significantly weaker than in the case of constant $\phi$, by a factor 0.83 as compared with 0.67. For $m_{\Phi}/H \gae 
0.5$, the value of $\xi^{1/2}$ is almost unaltered from the minimal D-term inflation model without RH sneutrino modification.

                From Figure \ref{si:b} we see that the reduction of the spectral index from the constant $\phi$ case is much less sensitive than $\xi^{1/2}$ to $m_{\Phi}/H$.  For $m_{\Phi}/H = 1.0$ we find that $n_{s} = 0.968$, which is still a significant suppression compared with the unmodified D-term inflation value, $n_{s} = 0.983$.

               In Figure \ref{rsi} we show the running spectral index, $\alpha = dn_{s}/d\ln k$. For small $m_{\Phi}/H$,
$\alpha$ tends towards a small negative value which is consistent with the constant $\phi$ value obtained from \eq{e14}, $\alpha = -0.00026$. As 
$m_{\Phi}/H$ increases, $\alpha$ increases to positive values, with $\alpha = 0.0024$ for $m_{\Phi}/H = 0.5$ and 
$\alpha = 0.008$ for $m_{\Phi}/H = 1.0$. These are much larger than and of opposite sign to the value in conventional hybrid inflation models, 
$\alpha = -1/N^2 = -0.00028$, and may be large enough to be observable in the future. The values of $m_{\Phi}/H$ for which $\alpha$ is larger than 0.001 are too large to significantly suppress the cosmic string contribution to the CMB, but they are consistent with a significant suppression of $n_{s}$ relative to the value in conventional D-term inflation and so could be important in non-minimal D-term inflation models without cosmic strings \cite{semilocal}. Thus a 
positive $\alpha >  0.001$ combined with $n_{s}$ smaller than 0.983 would be a signature of non-minimal D-term inflation with RH sneutrino modification.

\begin{figure} 
  \centering 
    \includegraphics[width=4.0in, angle = -90]{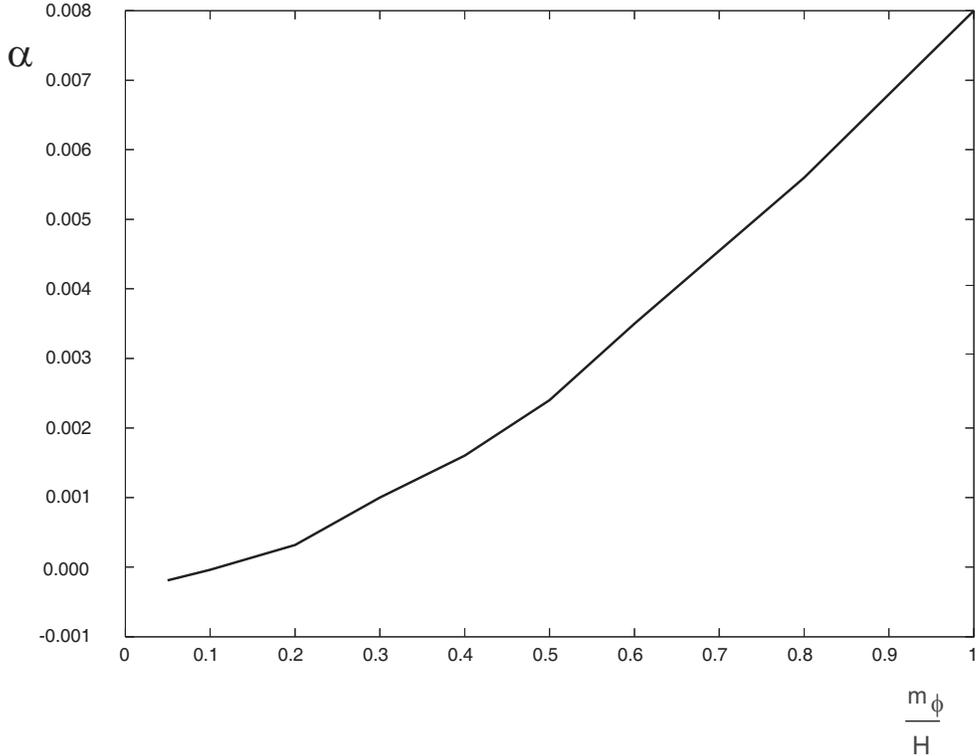} 
  \caption{Running spectral index $\alpha$ as a function of $m_{\Phi}/H$.  } 
  \label{rsi} 
\end{figure}

      For the results of the constant $\phi$ approximation to be reasonable we should restrict to  
$m_{\Phi}/H \lae 0.1$. In this case we need to check if it is still possible for the RH sneutrino to modify D-term inflation when $|\Phi|^2/M^2$ is small compared with 1.  For example, if we require $\gamma \approx 1$
in order to sufficiently modify $n_{s}$ and $\xi^{1/2}$                   
then \eq{e18} with $m_{\Phi}/H \lae 0.1$ implies that 
\be{x2}     \frac{|\Phi|^{2}}{M^{2}} =  \frac{3 H^{2} \gamma}{2 \left(c-1\right) m_{\Phi}^{2} N}
 \gae \frac{2.5}{\left(c-1\right)}      ~.\ee
$|\Phi|^{2}/M^{2}$ should be sufficiently small to keep SUGRA corrections to the RH sneutrino potential under control. For example, $c = 8$ and $m_{\Phi}/H \approx 0.1$ require that $|\Phi|^2/M^2 \approx 0.35$. Therefore relatively large values of $c$ are required. The RH sneutrino modification for a given $|\Phi|^2/M^2$ can be enhanced by the effect of additional RH sneutrino generations. In the idealized case where all three generations have the same mass and expectation value, the value of $|\Phi|^2/M^2$ will be smaller by a factor of 3 for a given $c$ and $m_{\Phi}/H$, thus allowing smaller values of $c$.

        Since $|\Phi|^2/M^2$ is generally not very small compared with 1, SUGRA corrections to the RH sneutrino potential could become significant. In fact, such corrections could help to enhance RH sneutrino modification. For example, if the leading order correction to $V(\phi)$ has a negative sign, so flattening $V(\phi)$ at large $\phi$, then it is possible that the rate of roll of the RH sneutrino could be smaller than in the case of a $\phi^2$ potential for a given value of $V(\phi)$. This would allow the 
constant $\phi$ approximation to hold even if the RH sneutrino mass was not very small compared with $H$.

\section{Conclusions}

      We have shown that including SUGRA corrections allows the RH sneutrino-modified variant of minimal D-term inflation to be consistent with CMB bounds on the spectral index even when the CMB contribution from cosmic strings is small. The value of $n_{s}$ is predicted to be low, less than 0.968 when $g \leq 1.0$ for a cosmic string contribution less than 5$\%$, while cosmic strings are expected to contribute at least 1$\%$ to the 
CMB power spectrum for $g \leq 1$. The combination of low spectral index plus greater than $1 \%$ percent CMB cosmic string contribution may be regarded as a prediction of the RH sneutrino-modified D-term inflation model.   

      We have anaylsed the model as a two-field inflation model using the $\delta N$ method. The main two-field inflation effect is from the time evolution of the classical RH sneutrino field, with quantum fluctuations of the RH sneutrino field having a negligible effect on the curvature perturbation. 

Two-field inflation effects strongly diminish the RH sneutrino suppression of the cosmic string tension once $m_{\Phi}/H \gae 0.2$, in which case the 
model will violate CMB bounds on cosmic strings. However, the effect on RH sneutrino suppression of the spectral index is smaller, which may be significant for D-term inflation models which evade cosmic string formation but still have a large spectral index. For $m_{\Phi}/H \lae 0.1$ it is a good approximation to treat the model as a single-field inflation model with constant $\phi$. 

          The running spectral index in the limit $m_{\Phi}/H < 0.1$ is negative and of the same magnitude as in conventional hybrid inflation models, $\alpha \approx -0.0002$. For larger $m_{\Phi}/H$, $\alpha$ increases to positive values due to the rolling of the $\Phi$ field. 
Values of 
$\alpha$ greater than 0.008 are found for $m_{\Phi}/H > 1.0$. For these values of $m_{\Phi}/H$ there is no significant suppression of the CMB contribution from cosmic strings. However, the spectral index is still reduced relative to conventional D-term inflation, so a large positive $\alpha$ could occur in non-minimal D-term inflation models without cosmic strings. Observation of a positive running spectral index together with 
a reduced spectral index may therefore provide a signature of non-minimal D-term inflation with RH sneutrino modification.     

  In conclusion, we have shown that RH sneutrino-modified D-term inflation provides a version of the original minimal D-term inflation model which is completely consistent with CMB observations. The model requires no suppression of couplings and no additional fields beyond those that already exist in the minimal D-term inflation model and the MSSM with neutrino masses. It therefore provides a predictive candidate for SUGRA inflation with all the advantages of conventional D-term inflation with respect to naturalness. RH sneutrino modification can also play a role in bringing non-minimal D-term inflation models without cosmic strings into agreement with the observed spectral index.

\section*{Acknowledgement}      This work was supported (in part) by the European Union through the Marie Curie Research and Training Network "UniverseNet" (MRTN-CT-2006-035863) and by STFC (PPARC) Grant PP/D000394/1.


\begin{thebibliography}{50}


\bibitem{eta} E.J.Copeland, A.R.Liddle, D.H.Lyth. E.D.Stewart and D.Wands, \pr{D49}{1994}{6410}.

\bibitem{fti} G.R.Dvali, Q.Shafi and R.K.Schaefer, \prl{73}{1994}{1886}. 

\bibitem{etahi}  A.D.Linde and A.Riotto, \pr{D56}{1997}{1841}. 

\Bibitem{dti}   E.Halyo, \pl{B387}{1996}{43}; P.Binetruy and G.Dvali, \pl{B388}{1996}{241}.

\bibitem{dtics} R.Jeannerot, \pr{D56}{1997}{6205}; 
R.Jeannerot, J.Rocher and M.Sakellariadou, \pr{D68}{2003}{103514}.

\bibitem{wmap} D.N.Spergel et al, {\it astro-ph/0603449}.


\bibitem{wyman}    
  L.~Pogosian, S.~H.~H.~Tye, I.~Wasserman and M.~Wyman,
  Phys.\ Rev.\  D {\bf 68} (2003) 023506
  [Erratum-ibid.\  D {\bf 73} (2006) 089904]
  [arXiv:hep-th/0304188]; 
  L.~Pogosian, M.~C.~Wyman and I.~Wasserman,
  [arXiv:astro-ph/0403268];
  M.~Wyman, L.~Pogosian and I.~Wasserman,
  Phys.\ Rev.\  D {\bf 72} (2005) 023513
  [Erratum-ibid.\  D {\bf 73} (2006) 089905]
  [arXiv:astro-ph/0503364].


\bibitem{endo} M.Endo, M.Kawasaki and T.Moroi, \pl{B569}{2003}{73}.

\bibitem{csb} 
  J.~Rocher and M.~Sakellariadou,
  JCAP {\bf 0503} (2005) 004
  [arXiv:hep-ph/0406120];
  J.~Rocher and M.~Sakellariadou,
  Phys.\ Rev.\ Lett.\  {\bf 94}, 011303 (2005)

\bibitem{csb2} J.~Rocher and M.~Sakellariadou, JCAP {\bf 0611}, 001 (2006).


\bibitem{battye}
  R.~A.~Battye, B.~Garbrecht and A.~Moss,
  JCAP {\bf 0609}, 007 (2006)
  [arXiv:astro-ph/0607339].

 
\bibitem{aur}   A.A.Fraisse,  {\it JCAP} {\bf 0703} (2007) 008.

\bibitem{bevis} N.Bevis, M.Hindmarsh, M.Kunz and J.Urrestilla, \pr{D75}{2007}{065015}.

\bibitem{bj}
  B.~Clauwens and R.~Jeannerot,
  arXiv:0709.2112 [hep-ph].

\bibitem{dlin1} C-M.Lin and J.McDonald, \pr{D74}{2006}{063510}.

\bibitem{osyok} O.Seto and J.Yokoyama, \pr{D73}{2006}{023508}. 

\bibitem{deltan} M.Sasaki and E.D.Stewart, \ptp{95}{1996}{71}; A.A.Starobinsky, \pl{B117}{1982}{175}; 
A.A.Starobinsky, {\it JETP Lett.} {\bf 42} (1985) 152; D.H.Lyth, K.A.Malik and M.Sasaki, {\it JCAP} 
{\bf 0505} (2005) 004.  

\bibitem{deltan2} F.Vernizzi and D.Wands, {\it JCAP} 
{\bf 0605} (2006) 019; D.H.Lyth and Y.Rodriguez, \prl{95}{2005}{121302}. 

\bibitem{seesaw}  M.Gell-Mann, P.Ramond and R.Slansky in {\it Supergravity}, eds. P.van Nieuwenhuizen and D.Freedman 
(North-Holland, Amsterdam, 1979) p.315; T.Yanagida in {\it Proc. of the Workshop on the Unified Theory and the Baryon Number in the Universe}, eds. O.Sawada and A.Sugamoto (KEK Report No.79-18, Tsukuba, 1979) p.95; R.N.Mohaparta and G.Senjanovic, \prl{44}{1980}{912}.


\bibitem{hilltop} L.Boubekeur and D.H.Lyth, {\it JCAP} {\bf 0507} (2005) 010.  


\bibitem{hilltop2} K.Kohri, C-M.Lin and D.Lyth, {\it hep-ph/0707.3826}. 


\Bibitem{lr} D.Lyth and A.Riotto, \prep{314}{1999}{1}.

\bibitem{tfi} C.Gordon, D.Wands, B.A.Bassett and R.Maartens, \pr{D63}{2001}{023506}. 

\bibitem{semilocal} J.Urrestilla, A.Achucarro and A.C.Davis, \prl{92}{2004}{251302}.




\end{thebibliography}
\end{document}